\documentclass[12pt]{article}

\columnwidth\textwidth

\begin{document}

\newcommand{\be}{\begin{equation}}
\newcommand{\ee}{\end{equation}}
\newcommand{\beann}{\begin{eqnarray*}}
\newcommand{\eeann}{\end{eqnarray*}}
\newcommand{\bea}{\begin{eqnarray}}
\newcommand{\eea}{\end{eqnarray}}
\newcommand{\nn}{\nonumber}
\newcommand{\ben}{\begin{enumerate}}
\newcommand{\een}{\end{enumerate}}
\newcommand{\lb}{\label}
\newtheorem{df}{Definition}
\newtheorem{thm}{Theorem}
\newtheorem{lem}{Lemma}
\newtheorem{prop}{Proposition}
\begin{titlepage}

\noindent
\begin{flushright}
 BUTP-00/13 \\
 Freiburg THEP-00/8
\end{flushright}
\vspace*{1cm}
\begin{center}
{\large\bf Embedding variables} \\[0.5cm]
{\large\bf in the canonical theory of gravitating shells} 

\vskip 1cm
{\bf Petr H\'{a}j\'{\i}\v{c}ek} 
\vskip 0.4cm
Institut f\"ur Theoretische Physik, \\
Universit\"at Bern, \\
Sidlerstrasse 5, 3012 Bern, Switzerland, \\
E-mail: hajicek@itp.unibe.ch, \\
Telephon: (41)(31) 631 86 25, FAX: 631 38 21.  
\vskip 0.7cm

{\bf Claus Kiefer} 
\vskip 0.4cm
Fakult\"{a}t f\"ur Physik,\\ Universit\"{a}t Freiburg, \\
Hermann-Herder-Strasse 3,
79104 Freiburg, Germany.
\vspace{1cm}

\nopagebreak[4]

\begin{abstract}
  A thin shell of light-like dust with its own gravitational field is studied
  in the special case of spherical symmetry. The action functional for this
  system due to Louko, Whiting, and Friedman is reduced to Kucha\v{r} form:
  the new variables are embeddings, their conjugate momenta, and Dirac
  observables. The concepts of background manifold and covariant gauge fixing,
  that underlie these variables, are reformulated in a way that implies the
  uniqueness and gauge invariance of the background manifold. The reduced
  dynamics describes motion on this background manifold.
\end{abstract}
\vskip 0.4cm
{\em PACS:} 0460


\end{center}

\end{titlepage}

\section{Introduction}
\label{sec:intro}
The phenomenon of gravitational collapse leads to serious problems in the
classical theory of gravity. The structure of the resulting singularities
contradict the foundations of the theory such as the equivalence principle.
Thus, the existence of singularity theorems \cite{H-E} may constitute a strong
motivation to address the quantisation of the gravitational field, in the hope
that such a framework can avoid the occurrence of singularities.

A prominent feature of the classical collapse is the existence of horizons
which appear sooner than the singularity. Such horizons not only imply that
the singularity is inevitable (which is, roughly, the content of the
singularity theorems), they also seem to prevent any object or information
from leaving the region of collapse and from coming back to the asymptotic
region. It is the existence of horizons that makes gravity so different and
the problem of collapse so difficult. On the other hand, the problem of
gravitational collapse is a very special one. For its solution, a complete
quantum theory of gravitation may be as little needed as the complete quantum
electrodynamics was needed for the first calculations of atomic spectra.

Motivated by these ideas, we consider the quantum theory of a spherically
symmetric thin shell and its gravitational field. This is, in fact, a quite
popular system. For example, it was used to study the motion of domain
walls in the early Universe \cite{guth}, of black-hole evaporation
\cite{K-W}, of quantum black holes \cite{berez}, and many others.
In \cite{haj1}, \cite{H-K-K} and \cite{honnef}, gravitational
 collapse of such
a thin shell in its own gravitational field has been studied.
 The result was
analogous to what is known about the $s$-mode of the Coulomb problem. Two
aspects of this result were surprising. First, for low-mass shells, there
were stationary states with Sommerfeld spectrum and scattering states with
the wave packets describing the shell bouncing and re-expanding.
 The evolution was unitary.
Second, there was an analogue to the critical charge in the
 relativistic quantum
mechanics of atoms.  The role of charge was played by the rest mass of the
shell (not to be confused with its total energy) and the critical value of the
rest mass was about one Planck mass.  As in the case of relativistic atoms,
the quantum-mechanical description breaks down for
 supercritical ``charges''.

To understand these results was difficult. Even the simple
 scattering of the
sub-critical shells admitted several different interpretations.
 There were two
problems.  First, the radial coordinate of the shell, which served as the
argument of the wave function, did not possess the status of a quantum
observable. The Coulomb-like potential prevented one from constructing a
position operator similar to the Newton-Wigner operator.
 Second, the model was
completely reduced to the physical degrees of freedom,
 which in this case is
just the radius of the shell.  However, the value of the radius is not as
informative in a black-hole spacetime as it is,
for example, in Minkowski spacetime:
points with the same value of radial coordinate can lie in different
asymptotically flat regions.  These regions are separated by horizons. The
tools that were at one's disposal in \cite{haj1}, \cite{H-K-K} and
\cite{honnef} did not allow to decide whether the shell created a horizon
and then, consequently, re-expanded behind this horizon into a different
asymptotically flat section, or whether 
it did not create any horizon and re-expanded
into the same region from which it collapsed.

In \cite{haj3}, two remedies have been proposed. The first is to work with a
null (light-like) shell. The classical dynamics of such a shell is equivalent
to that of free photons on flat two-dimensional spacetime (the ``charge'' is
zero). For such a system, there is a well-defined position operator
\cite{wightman}. Moreover, it admits a simple description of its asymptotic
states, unlike Coulomb scattering.

The second idea is that the equation of motion for $r(t)$, which has been
obtained from Einstein's equations, must in fact result from a reduction to
true degrees of freedom of an action that contains the shell as well as the
gravitational field. Indeed, in the spherically-symmetric case,
the gravitational degrees of
freedom consist only of the gauge and the dependent ones.
 It seems that the reduction
has been performed in such a way that the information about the geometry
of spacetime has been lost.  We shall, therefore, perform the reduction
explicitly in a careful way. This is the main purpose of this
paper.

In general, the reduction procedure consists of two steps:
 the choice of gauge
and the solution of constraints. There exists a particular form of the
gravitational action that is effectively reduced, but
which still contains some information
about the geometry of spacetime: the so-called {\em
  Kucha\v{r} decomposition} \cite{3way}, \cite{canada}. Kucha\v{r} variables
are neatly separated into pure gauge ones (so-called embeddings), dependent
ones that are conjugate to the embeddings, and physical degrees of freedom.
Some progress in understanding Kucha\v{r} decomposition has been achieved
recently \cite{paris}, \cite{H-K}:
 general existence of the decomposition has
been shown, and the crucial role of gauge choice in it has been recognised.
The nature of gauge choice in quantum gravity has also been elucidated. Two
important notions have been introduced: {\em background manifold} and
{\em covariant gauge fixing}. As a matter of fact, the present paper is the
first practical application of these concepts.
It deals with the {\em classical} canonical analysis
of the null-dust shell. Its main result is the {\em explicit construction}
of the Kucha\v{r} decomposition. This then serves as the starting
point for the quantum theory of the shell, which will be presented
in a separate paper. 

The plan of the paper is as follows. Sect.~2 explains
the notions of background
manifold and covariant gauge fixing in a new and clear form.
This enables us
to show the gauge invariance, the uniqueness
and some additional structures of
the background manifold, which will be necessary
for the interpretation of the
shell quantum mechanics. Some important points of \cite{H-K} are then
summarised in the new language.
Sect.~3 describes the solutions of Einstein's
equations containing the shell. After fixing the gauge, these solutions are
reformulated as a set of parameter-dependent metric fields and shell
trajectories on the background manifold. The parameters distinguish the
physically different solutions and will 
play the role of the physical variables in the
Kucha\v{r} decomposition. The representative metric fields and shell
trajectories will be used to define the transformation from the ADM to
Kucha\v{r} variables on the constraint hyper-surface. This transformation is
performed in Sect.~4. Starting point is a (non-reduced) Hamiltonian action
principle \cite{L-W-F} for the spherically-symmetric shell and its
gravitational field. In Sect.~4, an extension of the results to a whole
neighbourhood of the constraint hyper-surface is performed
using the methods and
theorems of \cite{H-K}. The final action, the variables in it and some
discussion can be found in Sect.~5.

\section{Background manifold \\ and covariant gauge fixing}
In this section we introduce the two basic notions
of background manifold and covariant gauge fixing, restricting ourselves,
for the
sake of simplicity, to vacuum general relativity. The language is slightly
different from that used in \cite{H-K}
 so that we can prove more results; we
also summarise some points that are important for the paper.

Let $\mathcal M$ be a four-manifold that admits a Lorentzian metric field $g$
such that $({\mathcal M},g)$ is a globally hyperbolic spacetime. Dynamically
maximal solutions of Einstein's equations are always of this form
\cite{G-Ch}. Then according to a theorem of Geroch \cite{G}, ${\mathcal M} =
\Sigma \times {\mathbf R}$, where $\Sigma$ is an initial data
manifold. Topological sectors of general relativity are uniquely associated
with different three-manifolds $\Sigma$. $\mathcal M$ is called {\em
  background manifold}. In this way, each topological sector determines a {\em
  unique} background manifold $\mathcal M$.

All diffeomorphisms of $\mathcal M$ form the group Diff$\mathcal M$; this is
considered as the ``gauge group'' of general relativity.
 Observe that the group
depends on the topological sector chosen, i.e., on $\mathcal M$,
 and that
the manifold structure of $\mathcal M$ itself is {\em gauge invariant}. Single
points of $\mathcal M$ are, however, not
gauge invariant, since they are being pushed around by Diff$\mathcal
M$. In some important cases, not Diff$\mathcal M$ but some of its subgroups
play the role of gauge group. For example, in the case of asymptotically
flat space-times, only those diffeomorphisms are considered as gauge
transformations that become sufficiently quickly trivial at infinity. In
general, in such cases, $\mathcal M$ is equipped with some 
gauge-invariant structure in addition to the naked manifold one.

Let $\varphi \in$ Diff$\mathcal M$ and let $g$ be a Lorentzian metric on
$\mathcal M$. Then the inverse pull-back associated with $\varphi$ maps $g$
into another metric $g'$, $g' = (\varphi^{-1})_*g$. In this way, the group
Diff$\mathcal M$ acts on the space Riem$\mathcal M$ of all (suitably
restricted) Lorentzian metrics on $\mathcal M$. The action is not transitive
and so there are non-trivial orbits of the group in Riem$\mathcal M$. Such
orbits are called {\em geometries} and the quotient space Riem$\mathcal
M$/Diff$\mathcal M$ is the space of geometries. Let $\pi$ : Riem${\mathcal M}
\mapsto$ Riem$\mathcal M$/Diff$\mathcal M$ be
 the natural projection for the
quotient. One can equip
 the space of geometries with some additional structure, e.g., a
topology, starting from a structure of Riem$\mathcal M$ and using the
projection.

Suppose that we manage, at least for some open set $U \subset$ Riem$\mathcal
M$/Diff$\mathcal M$, to specify a section $\sigma$. This is a map,
\[
  \sigma : \mbox{Riem}{\mathcal M}/\mbox{Diff}{\mathcal M} \mapsto
  \mbox{Riem}{\mathcal M} 
\]
such that $\pi \circ \sigma =$ id. The meaning of such a section is that a
particular {\em representative metric} on $\mathcal M$
 is chosen for each geometry
in $U$. This is exactly what has been called {\em covariant gauge fixing} in
\cite{H-K}. Clearly, the transformation between two covariant gauge fixings is
not a single diffeomorphism, but an element of the Bergmann-Komar group
\cite{B-K}. 

Given a covariant gauge fixing $\sigma$ on $U$, one can use it to construct a
map from Riem$\mathcal M$/Diff${\mathcal M} \times$ Emb$(\Sigma,{\mathcal
  M})$, where Emb$(\Sigma,{\mathcal M})$ is the space of embeddings of the
initial data surface $\Sigma$ into $\mathcal M$, to the ADM phase space of
general relativity. The construction has been described in \cite{H-K} and it
goes, roughly, as follows. Let $\gamma \in U$ and let
$\sigma(\gamma)$ be the representative metric on $\mathcal M$. Let $X: \Sigma
\mapsto {\mathcal M}$ be an embedding. Then $\sigma(\gamma)$ determines the
first, $q_{kl}(x)$, and the second, $K_{kl}(x)$, fundamental forms of the
surface $X(\Sigma)$ in the spacetime $({\mathcal M}, \sigma(\gamma))$. The
corresponding point $\left(q_{kl}(x), \pi^{kl}(x)\right)$ of the ADM phase
space can be obtained in the well known way from $q_{kl}(x)$ and $K_{kl}(x)$.

In \cite{H-K}, this transformation has been restricted to give only the points
of the constraint surface $\Gamma$ of the ADM phase space; moreover, only
those points of $\Gamma$ have been selected,
where the evolved space-times do
not admit any isometry. Then the map from Riem$\mathcal
M$/Diff${\mathcal M} \times$ Emb$(\Sigma,{\mathcal M})$ to $\Gamma$ has been
shown to be invertible and extensible to a neighbourhood of Riem$\mathcal
M$/Diff${\mathcal M} \times$ Emb$(\Sigma,{\mathcal M})$ in the larger space
Riem$\mathcal M$/Diff${\mathcal M} \times T^*$Emb$(\Sigma,{\mathcal M})$,
which has then been mapped to a neighbourhood of $\Gamma$ in the ADM phase
space. Next, the Darboux-Weinstein theorem has been employed to prove some
nice symplectic properties of the map. These properties then make the map to a
general transformation of the ADM to the Kucha\v{r} variables.

This procedure will here be applied to the model of spherically-symmetric thin
gravitating shell in the subsequent sections. We shall  find
in the next section the set of
representative solutions for Einstein's equations for each physically distinct
situation of the shell because, as has been shown in \cite{H-K}, this
part of the section $\sigma$ suffices completely to construct the above map to
the constraint surface of our model.

\section{Einstein dynamics of the shell}
\label{sec:sol}
Any spherically-symmetric solution of Einstein's equations with a thin null
shell as the source has a simple structure. Inside the shell, the spacetime is
flat; outside the shell, it is isometric to a part of the Schwarzschild
spacetime of mass $M$. The two geometries must be stuck together along a
spherically-symmetric null hyper-surface so that the points with the same
values of the radial coordinate $R$ coincide.

All physically distinct solutions can be labeled by three parameters: $\eta\in
\{-1,+1\}$, distinguishing between the outgoing ($\eta = +1$) and in-going
($\eta = -1$) null surfaces; the asymptotic time of the surface, i.e., the
retarded time $u=T-R\in (-\infty,\infty)$ for $\eta = +1$, and the advanced
time $v=T+R\in (-\infty,\infty)$ for $\eta = -1$; and the mass $M\in
(0,\infty)$. An in-going shell creates a black-hole (event) horizon at $R=2M$
and ends up in the singularity at $R=0$. The outgoing shell starts from the
singularity at $R=0$ and emerges from a white-hole (particle) horizon at
$R=2M$.

We can write down the metric in the case $\eta = 1$ with the
help of retarded Eddington-Finkelstein coordinates $\tilde{U}$, $R$,
$\vartheta$ and $\varphi$. $\tilde{U}=u$ is the trajectory of the
shell, $\tilde{U}>u$ is a part of Minkowski spacetime,
\begin{equation}
  ds^2 = -d\tilde{U}^2 - 2d\tilde{U}dR + R^2d\Omega^2,
\label{1'1}
\end{equation}
and $\tilde{U}<u$ is a part of Schwarzschild spacetime,
\begin{equation}
  ds^2 = -\left(1 - \frac{2M}{R}\right)d\tilde{U}^2
    - 2d\tilde{U}dR + R^2d\Omega^2.
\label{1'2}
\end{equation}
Similarly, for $\eta=-1$, the advanced Eddington-Finkelstein
coordinates are $\tilde{V}$, $R$, $\vartheta$ and $\varphi$, and
$\tilde{V}=v$ is the shell. Inside the shell, $\tilde{V}<v$,
\begin{equation}
  ds^2 = -d\tilde{V}^2 + 2d\tilde{V}dR + R^2d\Omega^2,
\label{1'3}
\end{equation}
and outside the shell, $\tilde{V}>v$,
\begin{equation}
  ds^2 = -\left(1 - \frac{2M}{R}\right)d\tilde{V}^2 +
    2d\tilde{V}dR + R^2d\Omega^2.
\label{1'4}
\end{equation}

Let us denote the spacetime given by the triple of
parameters $\eta$, $M$ and $w$ by $(\eta,M,w)$, where $w=u$
for $\eta=1$ and $w=v$ for $\eta=-1$.

We observe that the two space-times $(\eta,M,w_1)$ and $(\eta,M,w_2)$ are
isometric, the isometry sending the point $(\tilde{U},R,\vartheta,\varphi)$
into $(\tilde{U}+w_2-w_1,R,\vartheta,\varphi)$. Hence, the geometries of the
solutions that differ only in the value of the parameter $w$ are equal. Yet,
the physical situations they represent are different; this is similar to the
motion of a free mass point in Minkowski spacetime. For each two different
trajectories, there is a Poincar\'{e} transformation that sends the first into
the second. Still, the two motions are physically different because they look
differently from one fixed inertial frame. For the shell, instead of an
inertial frame, we imagine that there is a
 {\em fixed} asymptotic family of
observers.  The group of these isometries is a symmetry group rather than a
gauge group. It can (and will) be employed to define a time evolution.

Another interesting isometry is the map ${\mathcal T} : (\eta,M,w_1) \mapsto
(-\eta,M,w_2)$ defined for $\eta=+1$ and arbitrary $w_1$ and $w_2$ by the
Eddigton-Finkelstein coordinates as follows:
\[
  {\mathcal T}(\tilde{U}_1,R_1,\vartheta_1,\varphi_1) =
  (\tilde{V}_2,R_2,\vartheta_2,\varphi_2)\ ,
\]
where
\[
  \tilde{V}_2 = -\tilde{U}_1 +w_1 + w_2,\quad R_2 = R_1,\quad
  \vartheta_2 = \vartheta_1,\quad \varphi_2 = \varphi_1\ .
\]
For $\eta = -1$, we just take the inverse of the above so
that ${\mathcal T}^2 = \mbox{id}$. $\mathcal T$ can be viewed as a time
reversal symmetry.

The Eddington-Finkelstein coordinates may be nicely adapted to the symmetry
and may simplify the metric, but they do not define a covariant gauge fixing.
Indeed, the identification of the points
$(\tilde{U}_1,R_1,\vartheta_1,\varphi_1)$ of the solution $(+1,M_1,w_1)$ with
the points $(\tilde{V}_2,R_2,\vartheta_2,\varphi_2)$ of $(-1,M_2,w_2)$
satisfying the relations $\tilde{V}_2 = \tilde{U}_1$, $R_2 = R_1$,
$\vartheta_2 = \vartheta_1$ and $\varphi_2 = \varphi_1$ will invert the time
orientation of the asymptotic observers, which is to stay gauge invariant. We
need, however, a covariant gauge fixing if we are to transform the action to
the Kucha\v{r} form. The rest of this section will be devoted to a choice of
gauge that will be convenient for this problem.

To start with, we have to specify the background manifold.
Our model comprises only
the spherically-symmetric part of general relativity with the shell. We shall,
therefore, admit only spherically-symmetric initial surfaces $\tilde{\Sigma}$
and only that subgroup of Diff$\tilde{\mathcal M}$ (where $\tilde{\mathcal M}
:= \tilde{\Sigma} \times {\mathbf R}$), the elements of which commute with the
rotations and are trivial at infinity. Let $\rho$, $\vartheta$ and $\varphi$
be coordinates on $\tilde{\Sigma}$ that are adapted to the symmetry and $\rho
\in [0,\infty)$, where $\rho = 0$ is the regular centre of symmetry; we assume
that $\tilde{\Sigma}$ is smooth at this centre. The shell is at $\rho =
{\mathbf r}$, and the infinity at $\rho = \infty$.

The coordinates $\vartheta$ and $\varphi$ are ignorable coordinates; in the
action, we can integrate over them so that they disappear and the effective
initial manifold $\Sigma$ is one-dimensional, diffeomorphic to ${\mathbf
  R}_+$, and the effective background manifold $\mathcal M$ is
two-dimensional, ${\mathbf R}_+ \times {\mathbf R}$. Our restricted gauge
group induces an effective gauge group, Diff$_{0,\infty}{\mathcal M}$, on
$\mathcal M$; it only contains diffeomorphisms that preserve the central
boundary as well as, pointwise, the infinity.

Let us choose coordinates $U$ and $V$ on $\mathcal M$ that satisfy the
following boundary conditions at the gauge-invariant boundaries of $\mathcal
M$: At the regular centre inside the shell, 
\[
  U = V\ ,
\]
at ${\mathcal I}^-$, $U = -\infty$ and $V \in (-\infty,\infty)$, at ${\mathcal
  I}^+$, $V = \infty$ and $U \in (-\infty,\infty)$, and at $i^0$, $U = -\infty$
and $V = \infty$. Otherwise, $U$ and $V$ are arbitrary.

Using these coordinates $U$ and $V$, one can define the representative metric
(see Sect.~2) by conditions on its components
 with respect to $U$ and $V$. We shall choose
them as follows.
\begin{enumerate}
\item $U$ and $V$ are double-null coordinates so that the representative line
  element takes the form
\begin{equation}
  ds^2  =  -A(U,V)dUdV + R^2(U,V)(d\vartheta^2 + \sin^2\vartheta d\varphi^2)\ .
\label{KRV}
\end{equation}
\item The representative metric is continuous at the shell.
\item For the outgoing shells, $U$ is the retarded time determined by the
  representative metric at $V = \infty$. Analogously, for the in-going shells,
  $V$ is the advanced time at $U = -\infty$.
\end{enumerate}

Such a metric is uniquely defined for any physical situation given by the
values of the parameters $\eta$, $M$, and $w$. This can be shown as follows.

Consider first the case $\eta=+1$. The Eddington-Finkelstein
coordinate $\tilde{U}$ satisfies already the conditions for $U$, so we
need only to find the function $V$. In the Minkowski part,
$U>u$, of the solution, the boundary conditions at the centre
lead uniquely to:
\begin{equation}
  A = 1\ ,\quad R = \frac{V-U}{2}\ .
\label{5'1}
\end{equation}

In the Schwarzschild part, $U<u$, of the solution, $V$ is an advanced
null coordinate, so it must be some function, $V = X(M,u,\tilde{V})$,
for each fixed $M$ and $u$, of the advanced Eddington-Finkelstein
coordinate $\tilde{V}$, which is defined by
\[
  \tilde{V} := U + 2R + 4M\ln\left|\frac{R}{2M} - 1\right|.
\]

The function $X$ is uniquely determined by the boundary
condition at the shell, requiring that $V$ be continuous:
\[ 
  X(M,u,\tilde{V}|_{U=u}) = (U + 2R)_{U=u}\ ,
\]
or,
\[
  X\left(M,u,u + 2R + 4M\ln\left|\frac{R}{2M} 
  - 1\right|\right) = u +2R.
\]
To solve this equation, we define
\[
  x := u + 2R + 4M\ln\left|\frac{R}{2M} - 1\right|\ ,
\]
calculate $R$ in terms of $M$, $u$, and $x$, and substitute
the result into the right-hand side:
\begin{equation}
  X(M,u,x) = u + 2R(M,u,x).
\label{6'2}
\end{equation}
A straightforward calculation yields
\begin{equation}
  R(M,u,x) = 2M
  \kappa\Biggl(\exp\Bigl(\frac{x-u}{4M}\Bigr)\Biggr)\ ,
\label{R()}
\end{equation}
where $\kappa$ is the well-known Kruskal function defined by
its inverse, 
\begin{equation}
  \kappa^{-1}(y) = (y-1)e^y\ ,
\label{6'1}
\end{equation} 
and $R>2M$ was used.
Eqs.\ (\ref{6'2}) and (\ref{R()}) yield 
\begin{equation} 
  V = u +
  4M\kappa\Biggl(\left(\frac{R}{2M} -
  1\right)\exp\left(\frac{U-u+2R}{4M}\right)\Biggr)\ . 
\label{7'1}
\end{equation}
A similar calculation for $R<2M$ leads to the same result. From this, it is
easy to calculate $R$, if we observe that
\[
  \frac{R}{2M} + \ln\left(\frac{R}{2M} - 1\right) =
  \ln\Biggl(\kappa^{-1}\left(\frac{R}{2M}\right)\Biggr)\ .
\]
Then Eq.\ (\ref{6'1}) implies that
\begin{equation} 
  R =
  2M\kappa\Biggl(\left(\frac{V-u}{4M} -
  1\right)\exp\left(\frac{V-U}{4M}\right)\Biggr)\ .
\label{7'2}
\end{equation}
This relation defines the desired transformation from
$(U,R,\vartheta,\varphi)$ to $(U,V,\vartheta,\varphi)$.

As the last step, we calculate the metric for $U<u$ in the new
coordinates. First, we differentiate the function $R$. The derivative
of (\ref{6'1}) determines the derivative of $\kappa$:
\be
  \kappa'(f) = \frac{1}{\kappa(f)e^{\kappa(f)}}\ ,
\label{kappa'}
\end{equation}
which holds for any $f$. Then,
\[
  dR = \frac{1}{2\kappa(f_+)e^{\kappa(f_+)}}\left[
  -\left(\frac{V-u}{4M}-1\right)\exp\left(\frac{V-U}{4M}\right)dU
  + \frac{V-u}{4M}\exp\left(\frac{V-U}{4M}\right)dV\right]
\]
with
\begin{equation}
  f_+ := \left(\frac{V-u}{4M}-1\right)\exp\left(\frac{V-U}{4M}\right).
\label{f}
\end{equation}
Now, (\ref{7'2}) implies that
\begin{equation}
  \frac{\kappa(f_+)-1}{\kappa(f_+)} = 1 - \frac{2M}{R}\ ,
\label{7'3}
\end{equation} 
and (\ref{6'1}) that $f = \Bigl(\kappa(f) -
1\Bigr)\exp\Bigl(\kappa(f)\Bigr)$. This leads to the relation
\[
  \frac{f_+}{\kappa(f_+)e^{\kappa(f_+)}} = 1 - \frac{2M}{R}\ ,
\]
and thus 
\[
  dR = -\frac{1}{2}\left(1 - \frac{2M}{R}\right)dU +
  \frac{1}{2\kappa(f_+)e^{\kappa(f_+)}}
  \frac{V-u}{4M}\exp\left(\frac{V-U}{4M}\right)dV\ .
\]
Substituting this into the metric (\ref{1'2}) results, finally, in
\be
  R = 2M\kappa(f_+),\quad A =
  \frac{1}{\kappa(f_+)e^{\kappa(f_+)}}\frac{V-u}{4M} 
  \exp\left(\frac{V-U}{4M}\right),
\label{KRV+}
\end{equation}
where $f_+$ is defined by (\ref{f}), cf.~(\ref{7'2}).  With these expressions,
it is easy to verify that $A$ and $R$ are continuous at the shell, as required.
We note that these expressions contain $u$ as well as $M$, which become
conjugate variables in the canonical formalism. This makes the transition to
the embedding variables non-trivial, and one must first look for this
transformation on the constraint surface.

In the case of in-going shells ($\eta=-1$) a completely analogous
procedure yields, for $V<v$, again (\ref{5'1}), and for $V>v$,
\be
  R = 2M\kappa(f_-),\quad A =
  \frac{1}{\kappa(f_-)e^{\kappa(f_-)}}\frac{v-U}{4M} 
  \exp\left(\frac{V-U}{4M}\right),
\label{KRV-}
\end{equation}
where
\[
  f_- := \left(\frac{v-U}{4M} - 1\right)\exp\left(\frac{V-U}{4M}\right).
\]
These expressions result from (\ref{KRV+}) by the substitution
$V-u\to v-U$.

As the result of the gauge fixing, the set of solutions $(\eta,M,w)$
can be written as a set of $(\eta,M,w)$-dependent metric fields
(\ref{KRV}) and a set of shell trajectories on a {\em fixed} background
manifold $\mathcal M$. Here, the corresponding functions $A$ and $R$
have the form
\begin{equation}
  A(\eta,M,w;U,V)\ ,\quad R(\eta,M,w;U,V)\ ,
\label{10'1}
\end{equation}
and the trajectory of the shell on the background manifold is simply $U=u$ for
$\eta = +1$ and $V=v$ for $\eta = -1$.

A key property of the background manifold is that it possesses a unique
asymptotic region with ${\mathcal I}^-$ defined by $U \rightarrow -\infty$ and
${\mathcal I}^+$ by $V \rightarrow +\infty$.  As the shell cannot escape the
background manifold, its reappearance at {\em an} asymptotic region must be
interpreted as the reappearance at the asymptotic region of $\mathcal M$.  In
this way, the background manifold is a tool to solve the problem of where the
shell reappears.

\section{Transformation to embedding variables}
\subsection{Canonical formalism}
\label{sec:canon}
The form of the canonical theory that is based on the embedding rather than
ADM-type variables has been studied and advocated by Kucha\v{r}. In the recent
paper \cite{H-K} a large step forward in this field has been achieved. The
embedding variables have been associated with background manifolds and gauge
fixings similar to what has been done in the previous section. The existence
of this transformation has been shown in the general case.

The resulting formalism inspires hopes that some unpleasant features of
the ADM variables can be removed. First, the ADM variables lead to
singular points in the physical configuration space (super-space
\cite{fischer,nico}) as well as at the constraint surface corresponding to
spaces or space-times with symmetries. Second, the symmetry of the ADM
theory itself is, on one hand, too large, containing all infinitesimal
surface deformations, including also those
transformations that do not result from
diffeomorphisms. On the other hand, it is too small because only
infinitesimal surface deformations and not finite group elements can
act on the whole phase space. The constraint surface that has been
constructed in \cite{H-K} has, however, the form of a fibre bundle,
which is a manifold (all points are regular), and the fibre group of
this bundle is the diffeomorphism group of the background manifold, so
it acts on the whole bundle.

As a Hamiltonian action principle that implies the dynamics of our system, we
take the action Eq.\ (2.6) of \cite{L-W-F} (see also \cite{K-W}). Let us
briefly summarise the relevant formulae. The spherically symmetric metric is
written in the form:
\[
  ds^2 = -N^2d\tau^2 + \Lambda^2(d\rho + N^{\rho}d\tau)^2 + R^2d\Omega^2\ ,
\]
and the shell is described by its radial coordinate $\rho = {\mathbf
  r}$. The action reads
\be
  S_0 = \int d\tau\left[{\mathbf p}\dot{\mathbf r} + \int
  d\rho(P_\Lambda\dot{\Lambda} + P_R\dot{R} - H_0)\right]\ ,
 \lb{S0}
\ee
and the Hamiltonian is
\[
  H_0 = N{\mathcal H} + N^\rho{\mathcal H}_\rho\ + N_\infty E_\infty\ ,
\] 
where $N_\infty := \lim_{\rho \rightarrow \infty} N^{\rho}(\rho)$,
 $E_\infty$ is the
ADM mass (see \cite{L-W-F}), $N$ and $N^\rho$ are the lapse and shift
functions, $\mathcal H$ and ${\mathcal H}_\rho$ are the constraints,
\begin{eqnarray}
 {\mathcal H} & = & \frac{\Lambda P_\Lambda^2}{2R^2} -
 \frac{P_\Lambda P_R}{R}  + 
 \frac{RR''}{\Lambda} - \frac{RR'\Lambda'}{\Lambda^2} 
 + \frac{R^{\prime 2}}{2\Lambda} - \frac{\Lambda}{2} + \frac{\eta{\mathbf
 p}}{\Lambda}\delta(\rho - {\mathbf r})\ ,
\label{LWF-H} \\
  {\mathcal H}_\rho & = & P_RR' - P_\Lambda'\Lambda - {\mathbf
 p}\delta(\rho - {\mathbf r})\ ,
\label{LWF-Hr}
\end{eqnarray}
and the prime (dot) denotes the derivative
with respect to $\rho$ ($\tau$).

The main topic of this paper is to transform the variables in the action
$S_0$.  This transformation will be split into two steps. The first step is a
{\em transformation} of the canonical coordinates $\mathbf r$, $\mathbf p$,
$\Lambda$, $P_\Lambda$, $R$, and $P_R$ {\em at the constraint surface}
$\Gamma$ that is defined by the constraints (\ref{LWF-H}) and (\ref{LWF-Hr}).
The new coordinates are $u$ and $p_u=-M$ for $\eta = +1$, $v$ and $p_v=-M$ for
$\eta = -1$, and the so-called embedding variables $U(\rho)$ and $V(\rho)$.

The second step is an {\em extension} of the functions $u$, $v$, $p_u$, $p_v$,
$U(\rho)$, $P_U(\rho)$, $V(\rho)$,
 and $P_V(\rho)$ {\em out of the constraint
surface}, where the functions $u$, $v$, $p_u$, $p_v$, $U(\rho)$, and $V(\rho)$
are defined by the above transformation, and $P_U(\rho)$, $P_V(\rho)$ by
$P_U(\rho)\vert_{\Gamma} = P_V(\rho)\vert_{\Gamma} = 0$.
 The extension must satisfy the condition that the
functions form a canonical chart in a neighbourhood of $\Gamma$. A proof that
such extension exists in general has been given in \cite{H-K}.

\subsection{Transformation functions at the constraint surface}
The constraint surface contains only points of the phase space that correspond
to initial data for {\em solutions} of Einstein's equations. Hence, we can
assume that the metric (\ref{KRV}) is a spherically-symmetric solution with a
shell, and so the functions $A(\eta,M,w;U,V)$ and $R(\eta,M,w;U,V)$ are those
written down in the previous section, Eqs.\ (\ref{5'1}), (\ref{KRV+}) and
(\ref{KRV-}). According to Sect.~2, if such a metric is given, then, for each
embedding, a unique first and second fundamental form can be calculated from
it, and so the map from the embeddings to the ADM variables $q_{kl}(x)$ and
$\pi^{kl}(x)$ can be constructed.

A very important point is to specify the family of embeddings that will be
used throughout the paper. The embeddings are given by
\[
  U = U(\rho),\quad V = V(\rho)\ .
\]
These functions have to satisfy several conditions.
\begin{enumerate}
\item As $\Sigma$ is spacelike, $U$ and $V$ are null and increasing towards
  the future, we must have $U' < 0$ and $V' > 0$ everywhere.
\item At the regular centre, the four-metric is flat and the three-metric is
  to be smooth. This implies $U'(0) = -V'(0)$  in addition to the condition
  $U(0) = V(0)$. This follows from $T'(U(0),V(0))=0$ and means that
  $\Sigma$ must run parallel to $T =$ const. in order to avoid
  conical singularities.
\item At infinity, the four-metric is the Schwarzschild metric. We require
  that the embedding approaches the Schwarzschild-time-constant surfaces $T =$
  const, and that $\rho$ becomes the Schwarzschild curvature coordinate $R$
  asymptotically. More precisely, the behaviour of the Schwarzschild
  coordinates $T$ and $R$ along each embedding $U(\rho),V(\rho)$ must satisfy
  \bea T(\rho) & = & T_\infty + O(\rho^{-1}),
  \label{Tinf} \\
  R(\rho) & = & \rho + O(\rho^{-1})\ .
  \label{Rinf}
  \end{eqnarray}
  The asymptotic coordinate $T_\infty$ is a gauge-invariant quantity and it
  possesses the status of an observable. 
\item At the shell ($\rho = {\mathbf r}$) we require the functions $U(\rho)$
  and $V(\rho)$ to be $C^\infty$. In fact, as the four-metric is continuous in
  the coordinates $U$ and $V$, but not smooth, only the $C^1$-part of this
  condition is gauge invariant. Jumps in all higher derivatives 
  are gauge dependent, but the condition will simplify equations.
\end{enumerate}
 
We suppose further that there is a whole foliation of the solution space-times.
Any foliation can be considered as a one-parameter family of embeddings:
\[
  U = U(\tau,\rho),\quad  V = V(\tau,\rho),
\]
the parameter being $\tau$. The metric (\ref{KRV}) reads, in terms of the
coordinates $\tau$, $\rho$, $\vartheta$ and $\varphi$:
\[
  ds^2 = -A\dot{U}\dot{V}d\tau^2 - A(\dot{U}V'+\dot{V}U')d\tau d\rho -
  AU'V'd\rho^2 + R^2d\Omega^2.
\]
{}From this metric, we can read off the values of the  variables $\Lambda$,
$R$, $N^{\rho}$ and $N$ immediately: 
\be 
  \Lambda = \sqrt{-A(o,U,V)U'V'},\quad R = R(o,U,V),
\label{lambda-R}
\end{equation}
where $o$ symbolises the observables ($w$ and $M$, respectively), and
\[
  N = -\frac{\dot{U}V' - \dot{V}U'}{2U'V'}\sqrt{-AU'V'},
    \quad N^{\rho} =
  \frac{\dot{U}V' + \dot{V}U'}{2U'V'}\ . 
\]
The expression $\dot{U}V' - \dot{V}U'$ is the Jacobian of the transformation
from $\tau$ and $\rho$ to $U$ and $V$, and we assume it to be positive.

To calculate the gravitational momenta, we can use the canonical equations
that follow when the  action $S_0$ is varied with respect to $P_\Lambda$
and $P_R$: 
\beann
  P_\Lambda & = & -\frac{R}{N}(\dot{R} - N^{\rho}R')\ , \\
  P_R       & = & -\frac{\Lambda}{N}(\dot{R} - N^{\rho}R') -
  \frac{R}{N}\left(\dot{\Lambda} - (N^\rho\Lambda)'\right)\ .
\eeann
Substituting for $R$, $\Lambda$, $N$ and $N^{\rho}$ gives
\beann
  \dot{\Lambda} - (N^{\rho}\Lambda)' & = &
    \frac{1}{2\Lambda}\frac{\dot{U}V' -
    \dot{V}U'}{2U'V'} \left(-A_UU^{\prime 2}V' + A_VU'V^{\prime 2} - AU''V' +
    AU'V''\right) \\
  \dot{R} - N^{\rho}R' & = & \frac{\dot{U}V' - \dot{V}U'}{2U'V'} \left(R_UU' -
    R_VV'\right),
\eeann
so that, finally,
\bea
  P_\Lambda & = & \frac{R}{\sqrt{-AU'V'}}\left(R_UU' - R_VV'\right), 
\label{plambda} \\
  P_R & = & R_UU' - R_VV' + \frac{RA_U}{2A}U' - \frac{RA_V}{2A}V' +
  \frac{R}{2}\frac{U''}{U'} - \frac{R}{2}\frac{V''}{V'}\ .
\label{pr}
\eea  
Here, the indices $U$ and $V$ denote the partial derivatives with respect to
$U$ and $V$.

Eqs.\ (\ref{lambda-R}), (\ref{plambda}), and (\ref{pr}) are the
transformation equations expressing the variables $\Lambda$, $R$, $P_\Lambda$
and $P_R$ in terms of the new variables at the constraint surface. The
functions $A$ and $R$ are given by (\ref{5'1}), (\ref{KRV+}), and
(\ref{KRV-}). 

We now turn to the remaining  variables $\eta$, $\mathbf r$ and $\mathbf
p$. We let $\eta$ unchanged; in fact, we shall consider the  action as
two different actions, one for each value of $\eta$.
The variable $\mathbf r$ is related to our new variables $u$, $v$, $M$,
$U(\rho)$ and $V(\rho)$ in a different way for each value of $\eta$. If $\eta
= +1$, then $\mathbf r$ is determined by the equation $U({\mathbf r}) = u$.
This is an equation with exactly one solution if $u$ satisfies the condition
$u < U(0)$ because $U(\rho)$ is a monotonous function with the range
$\Bigl(-\infty,U(0)\Bigr)$. For the differentials of the variables $U(\rho)$,
$\mathbf r$ and $u$, we obtain the relation:
\be
  d{\mathbf r} = \frac{du - dU({\mathbf r})}{U'({\mathbf r})}\ .
 \lb{drU}
\ee

Similarly, if $\eta = -1$, then ${\mathbf r}$ is defined by $V({\mathbf r}) =
v$ for $v > V(0)$, and the relation between the differentials takes the form:
\be
  d{\mathbf r} = \frac{dv - dV({\mathbf r})}{V'({\mathbf r})}\ .
\lb{drV}
\ee

The  variable $\mathbf p$ does not seem to be determined completely in
\cite{L-W-F} because equation (2.5a) of Ref.\ \cite{L-W-F}, which is
the only equation that could serve this purpose, does not make sense in the
limit $m\rightarrow 0$ of null shells. However, the  constraint equations
lead to some expressions for $\mathbf p$; these 
determine $\mathbf p$ only
at the constraint surface, but this is, in fact, all we need. Let
us, therefore, turn to the constraint equations.

\subsection{The constraints}
The constraint functions (\ref{LWF-H}) and (\ref{LWF-Hr}) contain finite
parts, which are obtained for $\rho \neq {\mathbf r}$ and in the limits $\rho
\rightarrow {\mathbf r}\pm$, and $\delta$-function parts. The
$\delta$-function parts can be rewritten as equations for finite quantities,
if one collects all terms with $\delta$-function and sets the coefficient
equal to zero.

{}From the boundary conditions at the shell and Eqs.\ (\ref{lambda-R}),
(\ref{plambda}) and (\ref{pr}), it follows that the functions $\Lambda(\rho)$
and $R(\rho)$ are continuous, whereas $\Lambda'(\rho)$, $R'(\rho)$,
$P_\Lambda(\rho)$ and $P_R(\rho)$ jump across the shell,
as the metric is not smooth. This implies in
turn that the $\delta$-function part of the constraints is equivalent to
\bea
  {\mathbf p} & = & -\eta R[R']\ ,
\label{pH} \\
  {\mathbf p} & = & -\Lambda[P_\Lambda]\ ,
\label{pHr}
\eea
where the symbol $[g] := g_+ - g_-$ denotes the jump of the quantity $g$
across the shell.

Let us calculate the jumps. We have
\[
  [R'] = [R_U]U' + [R_V]V'\ .
\]
For $\eta = +1$, we have to use (\ref{5'1}) and (\ref{KRV+}) and to
replace the limits $\rho \rightarrow {\mathbf r}\pm$ by $U \rightarrow u\pm$.
We obtain immediately from (\ref{5'1}) that
\[
  R_{U-} = -\frac{1}{2}\ ,\quad R_{V-} = \frac{1}{2}\ .
\]
Differentiating (\ref{KRV+}) with the help of formulae (\ref{kappa'})
and (\ref{f}) leads to, for $U < u$,
\[
  R_{U+} = -\frac{f_+}{2\kappa(f_+)e^{\kappa(f_+)}}\ ,\quad R_{V+} =
  \frac{\frac{V-u}{4M}\exp\frac{V-u}{4M}}{2\kappa(f_+)e^{\kappa(f_+)}}\ .
\]
Eq.\ (\ref{5'1}) and $\kappa(f_+)=R/2M$ imply for the limits that
\[
  \lim_{U\rightarrow u}\kappa(f_+) = \frac{V-u}{4M}\ ,
\]
so we have
\[
  R_{U+} = -\frac{1}{2} + \frac{2M}{V-u},\quad R_{V+} = \frac{1}{2}.
\]
Hence,
\[
  [R_U] = \frac{2M}{V-u},\quad [R_V] = 0\ .
\]
Similarly, for $\eta = -1$,
\[
  [R_U] = 0\ ,\quad [R_V] = -\frac{2M}{v-U}\ .
\]
There is also the relation
\[
  R|_{U=u} = \frac{V-u}{2}\ ,\quad R|_{V=v} = \frac{v-U}{2}\ ,
\]
and so (\ref{pH}) yields:
\bea
  \eta = +1:\quad {\mathbf p} & = & -MU'({\mathbf r})\ , 
\label{shellp+} \\
  \eta = -1:\quad {\mathbf p} & = & -MV'({\mathbf r})\ .
\label{shellp-}
\eea
For $P_\Lambda$, (\ref{plambda}) implies
\[
  \Lambda[P_\Lambda] = R[R_U]U' - R[R_V]V'\ ,
\]
and so (\ref{pHr}) gives the same result as (\ref{pH}).

Let us return to the finite part of the constraints (\ref{LWF-H}) and
(\ref{LWF-Hr}). If we substitute the above expressions
for $\Lambda$, $R$, $P_\Lambda$, and $P_R$, we obtain, after some
lengthy but straightforward calculation, for each $\rho \neq {\mathbf r}$:
\beann
  \lefteqn{{\mathcal H} = \frac{1}{2\Lambda}(4RR_{UV} + 4R_UR_V + A)U'V' +} \\
  && \frac{R}{A\Lambda}(AR_{UU} - A_UR_U)U^{\prime 2} +
  \frac{R}{A\Lambda}(AR_{VV} - A_VR_V)V^{\prime 2}\ , 
\eeann
and
\[ 
  {\mathcal H}_\rho = -\frac{R}{A}(AR_{UU} - A_UR_U)U^{\prime 2} +
  \frac{R}{A}(AR_{VV} - A_VR_V)V^{\prime 2}\ . 
\]
If $\mathcal H$ and ${\mathcal H}_\rho$ are zero for any embedding outside the
shell, that is for all possible $U'$ and $V'$, the coefficients of $U'V'$,
$U^{\prime 2}$ and $V^{\prime 2}$ must themselves vanish: \bea 4RR_{UV} +
4R_UR_V + A & = & 0,
\label{EUV} \\
  AR_{UU} - A_UR_U & = & 0,
\label{EUU} \\
  AR_{VV} - A_VR_V & = & 0.
\label{EVV}
\eea
These three equations are equivalent to the full set of
Einstein equations for any metric of the form (\ref{KRV}). Thus, our functions
$A$ and $R$ have to satisfy these equations. This is immediately clear for
(\ref{5'1}) which gives $A$ and $R$ inside the shell. A more tedious
calculation verifies the validity of
 (\ref{EUV}), (\ref{EUU}), and (\ref{EVV}) also
outside the shell, where $A$ and $R$ are given by (\ref{KRV+}) and
(\ref{KRV-}).

\subsection{Transformation of the Liouville form}
As it has been explained at the end of Sec.\ \ref{sec:canon}, the
transformation of the action (\ref{S0}) to the new variables will be
performed in two steps. The first step is restricted to the constraint
surface and forms the content of the present section. 

At the constraint surface, $H_0 = N_\infty E_\infty$ and the action
(\ref{S0}) becomes
\[
  S_0\vert_\Gamma = \int d\tau\left[{\mathbf p}\dot{\mathbf r} -
  N_\infty E_\infty + \int
  d\rho(P_\Lambda\dot{\Lambda} + P_R\dot{R})\right]\ .
\]
According to the discussion given in \cite{kuchS}, the ADM boundary
term $N_\infty E_\infty$ in the action can, after parametrisation at
the infinity, be written as $E_\infty \dot{T}_\infty$ and can be
considered as a part of a modified Liouville form. Let us denote this
form by $\Theta$:
\be
  \Theta = \int d\rho\left(P_{\Lambda}{d\Lambda} +P_R{dR}\right)
 +{\mathbf p}d{\mathbf r} - E_\infty dT_\infty.
\label{33}
\ee
As a result, the transformation of the action is nothing but the
transformation of the Liouville form $\Theta$. 

We expect that the terms remaining after the transformation do not depend on
any embeddings, because the pull-back of the symplectic form to the constraint
surface is degenerated exactly in the direction of the gauge variables
$U(\rho)$ and $V(\rho)$. As we shall see, the constraint surface $\Gamma$
consists of two components, $\Gamma^+$ and $\Gamma^-$, $\Gamma^+$ containing
all outgoing and $\Gamma^-$ all in-going shells. We split this form into three
terms for the in-going and outgoing part, respectively,
\[
  \Theta\vert_{\Gamma^+} = \Theta^+\vert_{\Gamma^+} + \Theta^-\vert_{\Gamma^+}
  + {\mathbf p}d{\mathbf r},
\]
where 
\[
  \Theta^+\vert_{\Gamma^+} = \int_{\mathbf r}^\infty d\rho\,(P_\Lambda
  d\Lambda + P_R dR) - MdT_\infty
\]
because $E_\infty = M$ at the constraint surface, and 
\[
  \Theta^-\vert_{\Gamma^+} = \int_0^{\mathbf r} d\rho\,(P_\Lambda
  d\Lambda + P_R dR)\ ,
\]
and similar expressions for $\Theta\vert_{\Gamma^-}$.
Let us first transform the
part $\Theta\vert_{\Gamma^+}$ of the Liouville form and make the ansatz
\be
\Theta^+\vert_{\Gamma^+}\equiv\int_{\mathbf r}^{\infty}
d\rho\ \vartheta\ -MdT_{\infty}\ ,
\ee
with
\be
\vartheta\equiv(fdU+gdV+h_i do^i)'+d\varphi\ ,\lb{vartheta}
\ee
where we have denoted the observables
$u$ and $M$ collectively by $o^i$ $(i=1,2)$.
This has to be compared with the corresponding part of (\ref{33}), where
the substitutions are made from (\ref{lambda-R}),
\bea
\frac{d\Lambda}{\Lambda} &=& \frac{A_U}{2A}dU
+\frac{A_V}{2A}dV+\frac{A_i}{2A}do^i+\frac{dU'}{2U'}
 +\frac{dV'}{2V'}\ , \\
dR &=& R_UdU+R_VdV+R_ido^i\ ,
\eea
and $P_\Lambda$ and $P_R$ are given by (\ref{plambda}) and
(\ref{pr}). 

It turns out to be convenient to make for the functions
in (\ref{vartheta}) the ansatz
\bea
f &=& \frac{RR_U}{2}\ln\left(-\frac{U'}{V'}\right)
 +F(U,V,o^i)\ ,\\
g &=& \frac{RR_V}{2}\ln\left(-\frac{U'}{V'}\right)+G(U,V,o^i)\ ,\\
h_i &=& \frac{RR_i}{2}\ln\left(-\frac{U'}{V'}\right)
 +H_i(U,V,o^i)\ ,\\
\varphi &=& RR_UU'-RR_VV'-\frac{R}{2}(R_UU'+R_VV')
 \ln\left(-\frac{U'}{V'}\right)\\
& & \; -FU'-GV'+\phi(U,V,o^i)\ .
\eea
The functions $F,G,H_i,\phi$ are then determined through
comparison with the coefficients of $dU,dV,dU',dV'$, and $do^i$.
This leads to the equations
\bea
F_V-G_U &=& \frac{R}{2A}(2AR_{UV}-A_UR_V-A_VR_U)\ , \lb{first} \\
H_{iU}-F_i &=& -\frac{R}{2A}(2AR_{iU}-A_iR_U-A_UR_i)\ ,\lb{second}\\
H_{iV}-G_i &=& \frac{R}{2A}(2AR_{iV}-A_iR_V-A_VR_i)\ , \lb{third}\\
\phi &=& 0\ .
\eea
We next calculate the right-hand side of these equations by using
the explicit expressions for $A$ and $R$ found in Sect.~2.
Outside the shell, these are the expressions (\ref{KRV+}).
It is convenient to introduce the abbreviations
\be
b=\frac{V-u}{4M},\quad a=\frac{U-u}{4M}\ .
\ee
One then has
\be
A=\frac{be^{b-a}}{\kappa e^{\kappa}},\quad
R=2M\kappa\ ,
\ee
where $\kappa$ is a function of $f_+=(b-1)e^{b-a}$, see (\ref{f}).
The following identity turns out to be useful:
\be
\frac{e^{b-a}}{e^{\kappa}}=\frac{\kappa-1}{b-1}\ .
\label{ident}
\ee
After some lengthy, but straightforward calculations one finds
\bea
F_V-G_U &=& -\frac{\kappa-1}{8b(b-1)}\ ,
\label{50} \\
H_{uU} &=& F_u-\frac{\kappa-1}{8b(b-1)}\ , 
\label{51} \\
H_{uV} &=& G_u+\frac{\kappa-1}{8b(b-1)}\ , 
\label{52} \\
H_{MU} &=& F_M-\frac{1}{2}-\frac{\kappa-1}{2(b-1)}\ , 
\label{53}\\
H_{MV} &=& G_M+\frac{\kappa-1}{2b(b-1)}a
 +\frac{\kappa^2-b^2}{2b(b-1)}\ .
\label{54}
\eea The freedom in the choice of solution to Eq.\ 
(\ref{first})--(\ref{third}) enable us to set $F\equiv 0$. From the first
equation one then gets 
\be G=\int_u^UdU\frac{\kappa-1}{8b(b-1)}=
-\frac{M(\kappa^2-b^2)}{4b(b-1)}\ , 
\lb{G} 
\ee 
where we have chosen the boundary condition that $G=0$ for $U=u$, i.e., at the
shell, and calculated the integral by the substitution $x=\kappa$.  One
recognises from (\ref{G}) that, at the shell, $G_M=0$ and $G_u=-1/8b$. With
this result for $G$, one can integrate Eqs.\ (\ref{51})--(\ref{54}) for $H_i$
and choose the integration constants such that
\bea
H_u &=& -G\ , \lb{Hu}\\
H_M &=& -\frac{1}{2}(U-u)-4bG \ ,\lb{HM} \eea having $H_i = 0$, $i = 1,2$, at
the shell. This then yields for the Liouville form outside the shell pulled
back to the constraint surface \be \Theta^+\vert_{\Gamma^+} =
(fdU+gdV+h_ido^i)\vert_{\mathbf r} ^{\infty}+ d\left(\int_{\mathbf
    r}^{\infty}d\rho\ \varphi\right) +d{\mathbf r}\varphi\vert_{\rho={\mathbf
    r}} - MdT_\infty\ . \lb{S0out} \ee

The fourth term on the right-hand side of
 (\ref{S0out}) is a total derivative and will be omitted,
since it does not contribute to the dynamics. Eqs.\ (\ref{drU}),
(\ref{shellp+}) and (\ref{shellp-}) lead to 
\[
  {\mathbf p}d{\mathbf r}=-M(du-dU)\ .
\]

Analogously, one finds for the part $\Theta^-\vert_{\Gamma^+}$ inside the
shell
\be
\Theta^-\vert_{\Gamma^+} = (kdU+ldV)\vert_0^{\mathbf r}
 +d\left(\int_0^{\mathbf r}d\rho\ \psi\right)-
 d{\mathbf r}\psi\vert_{\rho={\mathbf r}}\ , \lb{inside}
\ee
with
\bea
k &=& \frac{RR_U}{2}\ln\left(-\frac{U'}{V'}\right)\ ,\\
l &=& \frac{RR_V}{2}\ln\left(-\frac{U'}{V'}\right)\ ,\\
\psi &=& RR_UU'-RR_VV'
 -\frac{R}{2}(R_UU'+R_VV')\ln\left(-\frac{U'}{V'}\right)\ .
 \lb{S0in}
\eea

Compared to $f,g,h_i,\varphi$, there are no terms analogous to $G,F$, and
$H_i$, since the classical solutions (\ref{5'1}) inside the shell lead to a
vanishing right-hand side of (\ref{first})-- (\ref{third}).  Because of the
boundary condition $U'(0)=-V'(0)$, the functions $k$ and $l$ vanish at the
centre. The third term on the right-hand side of
 (\ref{inside}) is again a total derivative and will
be neglected.

One has therefore only potential contributions at the
{\em shell} and at {\em infinity}. We shall consider first the contribution
from the shell. Since there $F=G=H_u=H_M=0$, one has to calculate
\bea
& & (kdU+ldV){}\vert_{\rho={\mathbf r}}
-d{\mathbf r}\psi{}\vert_{\rho={\mathbf r}}
-(fdU+gdV+h_udu-h_MdM){}\vert_{\rho={\mathbf r}}
\nonumber\\
 & & \; +d{\mathbf r}\varphi\vert_{\rho={\mathbf r}}
 -M(du-dU)\ .
\eea
Using (\ref{5'1}) and (\ref{KRV+}), one arrives at the following jump
conditions at the shell:
\be
[RR_U]=M,\; [RR_V]=0,\;
[RR_u]=-M,\; [RR_M]=0\ .
\ee
Taking these into account, one recognises that {\em all} terms
on the dust shell {\em cancel}.
As we shall now demonstrate, the only non-vanishing terms are 
originating from {\em infinity}.
\begin{eqnarray}
 \lefteqn{\Theta\vert_{\Gamma^+} = \lim_{\rho\rightarrow\infty}
  \left[H_ido^i + FdU +
   GdV + \frac{}{}\right.} \nn \\
  && \left.\ln\left(\frac{-U'}{V'}\right)\left(\frac{RR_U}{2}dU +
     \frac{RR_V}{2}dV + \frac{RR_i}{2}do^i\right)\right] - MdT_\infty\ ,
\label{assL}
\end{eqnarray} 
where the function $F=0$, and $G$, $H_u$ and $H_M$ are given by 
(\ref{G}), (\ref{Hu}), and (\ref{HM}), respectively.
 The limit (\ref{assL}) is determined by
the boundary conditions 3 of Sec.\ 4.1, cf. (\ref{Tinf})
and (\ref{Rinf}).

Eqs.\ (\ref{Tinf}) and (\ref{Rinf}) determine the expansions of $U(\rho)$ and
$V(\rho)$ uniquely. Indeed, for $\eta =+1$, $U$ near the space-like infinity
coincides with the Edington-Finkelstein retarded coordinate (see Sect.~3)
 and so is given in terms of $T$ and $R$ by
\[
  U = T - R - 2M\ln\left(\frac{R}{2M} -1\right).
\]
Then,
\be
  U(\rho) = -\rho - 2M\ln\left(\frac{\rho}{2M}\right) + T_\infty +
  O(\rho^{-1}). 
\label{Uinf}
\end{equation}
The presence of the logarithmic term is due to the long range of the
gravitational potential. Thus, the first diverging term is universal, the
second depends on the observable $M$, and the asymptotic coordinate $T_\infty$
of the embedding appears only at the third position.

The asymptotic expansion of the function $V(\rho)$ can be determined from 
(\ref{7'1}). We have first to get rid of $\kappa$:
\be
  (V-u-4M)\exp\frac{V}{4M} = 2(R-2M)\exp\left(\frac{U}{4M} +
  \frac{R}{2M}\right). 
\label{Vinf1}
\end{equation}
Then we substitute the expansions (\ref{Rinf}) and (\ref{Uinf}) into the
right-hand side of (\ref{Vinf1}):
\bea
  \lefteqn{(V-u-4M)\exp\frac{V}{4M} = } \nn \\
  &&\left(\rho-2M + O(\rho^{-1})\right)\exp\left[\frac{1}{4M}\left(\rho -
      2M\ln\frac{\rho}{8M} + T_\infty\right) + O(\rho^{-1})\right]. 
\label{Vinf2}
\end{eqnarray}
Let us remove the singular part in the exponent by setting
\[
  V(\rho) = \rho - 2M\ln\frac{\rho}{8M} + T_\infty + V_1(\rho)\ .
\]
Eq.\ (\ref{Vinf2}) then becomes
\bea
  \lefteqn{(1-2M\rho^{-1})\exp O(\rho^{-1}) = } \nn \\
  &&\left[1 - 2M\rho^{-1}\ln\frac{\rho}{8M} + (T_\infty - u -
      4M)\rho^{-1} + 
    V_1(\rho)\rho^{-1}\right] \exp\frac{V_1(\rho)}{4M}\ . 
\label{Vinf3}
\end{eqnarray}
Taking the limits $\rho\rightarrow\infty$ of both sides of (\ref{Vinf3}),
we obtain
\[
  \lim_{\rho\rightarrow\infty}\left(1 +
  \frac{V_1(\rho)}{\rho}\right)\exp\frac{V_1(\rho)}{4M} = 1.
\]
It follows that
\[
  \lim_{\rho\rightarrow\infty} V_1(\rho) = 0\ .
\]
Eq.\ (\ref{Vinf3}) implies then also that
\[
  V_1(\rho) = O\left(\frac{\ln\rho}{\rho}\right).
\]
Hence, the expansion of the function $V(\rho)$ has the form:
\be
  V(\rho) = \rho - 2M\ln\frac{\rho}{8M} + T_\infty +
  O\left(\frac{\ln\rho}{\rho}\right);
\label{Vinf4}
\end{equation}
the asymptotic coordinate $T_\infty$ appears again only at the third position,
and this is the reason why the expansion must be carried so far.

Now, the expansion of all functions contained in $\Theta$ is a straightforward
matter. For $G$, we obtain from (\ref{G}):
\[
  G = -\frac{M}{4}\frac{4R^2 - (V-u)^2}{(V-u)^2 - 4M(V-u)}.
\]
Then, Eqs.\ (\ref{Rinf}), (\ref{Uinf}), and (\ref{Vinf4}) give
\[ 
  G = -\frac{3}{4}M - 4M^2\rho^{-1}\ln\frac{\rho}{8M} + M(2T_\infty - 2u
  -3M)\rho^{-1} + o(\rho^{-1})\ ,
\]
where $o(\rho^{-1})$ is defined by the property $\lim_{\rho \rightarrow
  \infty} \rho o(\rho^{-1}) = 0$, and 
\[
  H_u =-G= \frac{3}{4}M + 4M^2\rho^{-1}\ln\frac{\rho}{8M} - M(2T_\infty - 2u
  -3M)\rho^{-1} + O(\rho^{-2})\ ,
\]
as well as
\[ 
  H_M = \frac{5}{4}\rho + \frac{7}{2}M\ln\frac{\rho}{8M} + \frac{M}{2}(6 +
  4\ln2) - \frac{7}{4}(T_\infty - u) + O(\rho^{-1})\ .
\]
Eq.\ (\ref{Vinf4}) can be used to calculate $dV$:
\[
  dV = d\left(-2M\ln\frac{\rho}{8M}\right) + dT_\infty +
  O\left(\frac{\ln\rho}{\rho}\right)\ . 
\]
Then we obtain:
\[
  H_ido^i + FdU + GdV = M(dT_\infty - du) + dZ
  + O\left(\frac{\ln\rho}{\rho}\right),
\]
where
\[
  Z = -\frac{7}{4}M(T_\infty -u) +\frac{5}{4}M\rho + \frac{5}{2}M^2\ln\rho +
  \frac{1}{2}(4-13\ln 2)M^2 - \frac{5}{2}M^2\ln M.
\]
Similarly,
\[
  \frac{R}{2}\ln\left(\frac{-U'}{V'}\right) = 2M + O(\rho^{-2})\ .
\]
The derivatives $R_U$, $R_V$, $R_u$ and $R_M$ can be expanded if we calculate
them from (\ref{7'2}) using the identity (\ref{ident}):
\[
  R_U = -\frac{1}{2}\frac{R-2M}{R},\quad R_V =
  \frac{1}{2}\frac{R-2M}{R}\frac{V-u}{V-u-4M}\ , 
\]
\[
  R_u = -2M\frac{R-2M}{R}\frac{1}{V-u-4M}\ ,
\]
and
\[
  R_M = \frac{R}{M} - 2\frac{R-2M}{R}\left(\frac{1}{4M}\frac{V-U}{1 -
  \frac{4M}{V-u}} + \frac{U-u}{V-u-4M}\right).
\]
This gives:
\[
  \frac{R}{2}\ln\left(\frac{-U'}{V'}\right)(R_UdU + R_VdV +
  R_ido^i) = O\left(\frac{\ln\rho}{\rho}\right).
\]

Collecting all terms, we finally have:
\be
  \Theta\vert_{\Gamma^+} = -Mdu + dZ + O\left(\frac{\ln\rho}{\rho}\right).
\label{Thetainf0}
\end{equation}
The exact form $dZ$ can be omitted because it has no influence
 on the symplectic form and the
equations of motion.

The final result of this subsection can be formulated as follows. The
constraint surface $\Gamma$ consists of two components: $\Gamma^+$ for the
outgoing shells ($\eta = +1$), and $\Gamma^-$
 for the in-going shells ($\eta =
-1$). On $\Gamma^+$, we have the coordinates $M$, $u$, $U(\rho)$ and
$V(\rho)$, and the pull-back of the Liouville form to $\Gamma^+$
is
\[
  \Theta\vert_{\Gamma^+} = -Mdu.
\]
Thus, it is independent of $U(\rho)$ and $V(\rho)$, as expected.

In a completely analogous manner, the following result can be derived for the
$\eta = -1$ case:
\[
  \Theta\vert_{\Gamma^-} = -Mdv\ ,
\]
and our coordinates on $\Gamma^-$ are $M$, $v$, $U(\rho)$ and $V(\rho)$.

These results also show that the two Dirac observables $-M$ and $u$ (or $-M$
and $v$) form a conjugate pair. Indeed, the Poisson algebra of Dirac
observables is well-defined by the (degenerate) pull-back of the symplectic
form to the constraint surface.

\subsection{Extension to a neighbourhood of the constraint surface}

In the previous subsection, the constraint surface pull-back of the
Liouville form has been transformed to the Kucha\v{r} coordinates:
the embeddings $U(\rho)$ and $V(\rho)$ that represent pure gauges, and $p_u$
and $u$ (or $p_v$ and $v$) that are Dirac observables. The next task is to
extend these coordinates to a neighbourhood of the constraint surface $\Gamma
= \Gamma^+ \cup \Gamma^-$.

In \cite{H-K}, the proof has been given that such an extension exists if there
are no points with additional symmetry at $\Gamma$. For the action (\ref{S0}),
the space-time solutions are all spherically symmetric, but none of them
exhibits any other symmetry, be it discrete or continuous. We can reformulate
the result of \cite{H-K} in a way suitable for our purposes as follows.  There
is a neighbourhood $\Gamma^{\prime \pm}$ of $\Gamma^\pm$ in the phase space,
and functions 
\begin{equation}
  U, P_U, V, P_V, u, p_u,  
\label{76+}
\end{equation}
in $\Gamma^{\prime +}$ and 
\begin{equation}
  U, P_U, V, P_V, v, p_v,   
\label{76-}
\end{equation}
in $\Gamma^{\prime -}$ such that, at $\Gamma^\pm$,
\[
  P_U(\rho) = P_V(\rho) = 0,
\]
and $U(\rho)$, $V(\rho)$, $p_u$ and $u$ (or $p_v$ and $v$) coincide with our
coordinates there. The functions (\ref{76+}) and (\ref{76-}) form canonical
charts in $\Gamma^{\prime\pm}$. The transformation between the old variables
\begin{equation}
  R, P_R, \Lambda, P_\Lambda, {\mathbf r}, {\mathbf p}, 
\label{75}
\end{equation}
and the new ones Eq.\ (\ref{76+}) or (\ref{76-}) is smooth and invertible in
$\Gamma^{\prime\pm}$. At the constraint surface $\Gamma^\pm$, the
transformation is given by Eqs.\ (\ref{lambda-R})--(\ref{pr}), (\ref{KRV+}),
(\ref{KRV-}), $U({\mathbf r}) = u$, $V({\mathbf r}) = v$, (\ref{shellp+}) and
(\ref{shellp-}). Outside the constraint surface, only the existence of the
transformation has been shown so we do not know its form.

Using this result, we can write the transformed action $S^\pm$ in
$\Gamma^{\prime  \pm}$ as follows,
\beann
  S^+ & = &\int_{-\infty}^\infty d\tau \left[p_u\dot{u} + \int_0^\infty
    d\rho\left(P_U(\rho)\dot{U}(\rho) + \right. \right. \\
   && \left. \left. P_V(\rho)\dot{V}(\rho) - N_U(\rho)P_U(\rho) -
       N_V(\rho)P_V(\rho) \right) \right]\ ,
\eeann 
and 
\beann
  S^- & = &\int_{-\infty}^\infty d\tau \left[p_v\dot{v} + \int_0^\infty
    d\rho\left(P_U(\rho)\dot{U}(\rho) + \right. \right. \\
   && \left. \left. P_V(\rho)\dot{V}(\rho) - N_U(\rho)P_U(\rho) -
       N_V(\rho)P_V(\rho) \right) \right]\ .
\eeann 

The two actions can be considered as the 
reduced form of just one action. Consider
the case $\eta =+1$ first. The dynamical trajectory of the shell is given by
the relation $u(\tau) =$ const, whereas $v(\tau)$ is arbitrary, depending on
the choice of the parameter $\tau$ (the only restriction is that $v(\tau)$ is
an increasing function). This information can be obtained from the extended
action: 
\beann
  S^+_{\mbox{\footnotesize ext}} & = &\int_{-\infty}^\infty d\tau
  \left[p_u\dot{u} + p_v\dot{v} - n_up_v +\int_0^\infty
    d\rho\left(P_U(\rho)\dot{U}(\rho) + \right. \right. \\
   && \left. \left. P_V(\rho)\dot{V}(\rho) - N_U(\rho)P_U(\rho) -
       N_V(\rho)P_V(\rho) \right) \right]\ ,
\eeann 
where $n_u$ is a Lagrange multiplier , $v(\tau)$ a pure gauge
 and $p_v$-dependent. Similarly for $\eta = -1$:
\beann
  S^-_{\mbox{\footnotesize ext}} & = &\int_{-\infty}^\infty d\tau
  \left[p_u\dot{u} + p_v\dot{v} - n_vp_u +\int_0^\infty
    d\rho\left(P_U(\rho)\dot{U}(\rho) + \right. \right. \\
   && \left. \left. P_V(\rho)\dot{V}(\rho) - N_U(\rho)P_U(\rho) -
       N_V(\rho)P_V(\rho) \right) \right]\ .
\eeann
One can set in $S^+_{\mbox{\footnotesize ext}}$, as $p_u = -M < 0$:
\[
  n_u = np_u
\]
and, similarly, in $S^-_{\mbox{\footnotesize ext}}$, 
\[
  n_v = np_v\ .
\]
Then, clearly, $S^+_{\mbox{\footnotesize ext}}$ and $S^-_{\mbox{\footnotesize
    ext}}$ are obtained by reducing the following action
\beann
  S & = &\int_{-\infty}^\infty d\tau
  \left[p_u\dot{u} + p_v\dot{v} - np_up_v +\int_0^\infty
    d\rho\left(P_U(\rho)\dot{U}(\rho) + \right. \right. \\
   && \left. \left. P_V(\rho)\dot{V}(\rho) - N_U(\rho)P_U(\rho) -
       N_V(\rho)P_V(\rho) \right) \right]\ .
\eeann
Indeed, the case $\eta =+1$ ($\eta = -1$) is obtained from the solution $p_v =
0$ ($p_u = 0$) of the constraint 
\[
  p_up_v = 0\ .
\]
 
The relation between the total energy $M$ and the two momenta $p_u$ and $p_v$
can now be written as follows:
\[
  M = -p_u - p_v.
\]

\section{Conclusions}
We have demonstrated that there is a transformation of variables
bringing the action (\ref{S0}) to the simple form of the so-called
{\em Kucha\v{r}   decomposition}
\begin{equation}
  S = \int d\tau\left(p_u\dot{u} + p_v\dot{v} - np_up_v\right)
  + \int d\tau\int_0^\infty d\rho(P_U\dot{U} + P_V\dot{V} - H)\ ,
\label{KD}
\end{equation}
where $H = N^UP_U + N^VP_V$; $n$, $N^U(\rho)$ and $N^V(\rho)$ are the new
Lagrange multipliers. 

The dependence of the new variables $P_U$ and $P_V$ on the old ones is not
known. This dependence would be needed for calculation of the spacetime
geometry associated with any solution given in terms of the new variables. We
know, however, that the new constraint equations, $P_U(\rho) = P_V(\rho) = 0$,
are mathematically equivalent to the old constraints, Eqs.\ (\ref{LWF-H}) and
(\ref{LWF-Hr}). One can, therefore, use the old constraints to calculate the
geometry from the true degrees of freedom along the hypersurfaces of some
foliation. The fact that two spacetimes obtained by this method using
different foliations are isometric is guaranteed by the closure of the algebra
of constraints \cite{teitel}.

The new phase space has non-trivial boundaries:
\begin{equation}
  p_u \leq 0,\quad p_v \leq 0\ ,
\label{77}
\end{equation}
\begin{equation}
 \frac{-u+v}{2} > 0\ ,
\label{78}
\end{equation}
\begin{equation}
  p_v = 0\ ,\quad U \in (-\infty,u)\ , \quad V > u -
  4p_u\kappa\left(-\exp\frac{u-U}{4p_u}\right)\ , 
\label{79}
\end{equation}
and 
\begin{equation}
  p_u = 0\ ,\quad V \in (v,\infty)\ , \quad U < v +
  4p_v\kappa\left(-\exp\frac{V-v}{4p_v}\right)\ . 
\label{80}
\end{equation}
The boundaries defined by inequalities (\ref{78})--(\ref{80}) are due
to the singularity.

The two dynamical systems defined by the actions (\ref{S0}) and
(\ref{KD}) are equivalent: each maximal dynamical trajectory of the
first, if transformed to the new variables, give a maximal dynamical
trajectory of the second and vice versa.

The variables $u$, $v$, $p_u$ and $p_v$ span the effective phase space of the
shell. They contain all true degrees of freedom of the system. One can observe
that the corresponding part of the action (\ref{KD}) coincides with the action
for free motion of a zero-rest-mass spherically-symmetric (light) shell in
flat spacetime if one replaces the inequality (\ref{78}) by
\[
   \frac{-u+v}{2} \geq 0\ .
\]
Such a dynamics is complete if the singularity at the value zero of the radius
of the shell, $(-u+v)/2$, can be considered as a harmless caustic so that the
light can re-expand after passing through it. It might, therefore, seem also
possible to extend the phase space of the gravitating shell in the same way so
that the in-going and the out-going sectors are merged together into one
bouncing solution.

However, such a formal extension of the dynamics (\ref{KD}) is not
adequate. The
physical meaning of any solution written in terms of new variables (\ref{76+})
or (\ref{76-}) is given by measurable quantities of geometrical or physical
nature such as the curvature of spacetime or the density of matter. These
observables must be expressed as functions on the phase space. They can of
course be transformed between the phase spaces of the two systems (\ref{S0})
and (\ref{KD}). They cannot be left out from any complete description of a
system, though they are often included only tacitly: an action alone does not
define a system. This holds just as well for the action (\ref{S0}) as for
(\ref{KD}).

Let us consider these observables. The expression for the
stress-energy tensor of the shell written down in \cite{L-W-F} implies
that the density of matter diverges at ${\mathbf r} = 0$; this
corresponds to $(-u+v)/2 = 0$ in terms of the new variables. Eqs.\
(\ref{KRV+}) and (\ref{KRV-}) can be used to show that the curvature
of the solution spacetime diverges at the boundary defined by Eq.\
(\ref{79}) for $p_v = 0$ and by Eq.\ (\ref{80}) for $p_u = 0$. It
follows that the observable quantities at and near the ``caustic'' are
badly singular and that there is no sensible extension of the dynamics
defined by action (\ref{KD}) to it, let alone through it. This confirms the
more or less obvious fact that no measurable property (such as the
singularity) can be changed by a transformation of variables.

The action for the null dust shell is now written in a form which can
be taken as the starting point for quantisation. Surprisingly,
it will turn out that the quantum theory is, in fact, singularity-free.
 This will be done in
a separate paper \cite{PH}.

\subsection*{Acknowledgments}
Helpful discussions with K.~V.~Kucha\v{r} and L.~Lusanna are acknowledged. The
authors wish to thank I.~Kouletsis for checking
some of the equations. This work was
supported by the Swiss National Science Foundation and the Tomalla Foundation
Z\"{u}rich. C.K. thanks the University of Bern for its kind hospitality, while
part of this work was done; for the same reason, P.H. thanks the University
of Florence.

\end{document}